\documentclass{ws-procs11x85}
\usepackage{ws-procs-thm}           

\usepackage[utf8]{inputenc} 
\usepackage[T1]{fontenc}    
\usepackage{hyperref}       
\usepackage{url}            
\usepackage{booktabs}       
\usepackage{amsfonts}       
\usepackage{nicefrac}       
\usepackage{microtype}      
\usepackage{xcolor}         
\usepackage{float}

\begin{document}
\title{Acoustic-Linguistic Features for Modeling Neurological Task Score in Alzheimer's}
\author{Saurav K. Aryal$^\dag$ Howard Prioleau and Legand Burge}

\address{EECS, Howard University,\\
Washington, DC 20059, USA\\
$^\dag$E-mail: saurav.aryal@howard.edu\\
https://howard.edu/}

\begin{abstract}
  The average life expectancy is increasing globally due to advancements in medical technology, preventive health care, and a growing emphasis on gerontological health. Therefore, developing technologies that detect and track aging-associated disease in cognitive function among older adult populations is imperative. In particular, research related to automatic detection and evaluation of Alzheimer's disease (AD) is critical given the disease's prevalence and the cost of current methods. As AD impacts the acoustics of speech and vocabulary, natural language processing and machine learning provide promising techniques for reliably detecting AD. We compare and contrast the performance of ten linear regression models for predicting Mini-Mental Status Exam scores on the ADReSS challenge dataset. We extracted 13000+ handcrafted and learned features that capture linguistic and acoustic phenomena. Using a subset of 54 top features selected by two methods: (1) recursive elimination and (2) correlation scores, we outperform a state-of-the-art baseline for the same task. Upon scoring and evaluating the statistical significance of each of the selected subset of features for each model, we find that, for the given task, handcrafted linguistic features are more significant than acoustic and learned features.
\end{abstract}

\section{Introduction}

People are living longer due to advancements in medical technology, preventive health care, and a growing emphasis on gerontological health. The Administration for Community Living estimates that by 2020, 77 million people living in the United States will be 60 years of age or older. Hence, developing technologies that detect and track aging-associated disease in cognitive function among older adult populations is imperative.

For decades scientists have examined the association between psychological well-being and cognition. In prior research, gerontologists have identified a significant relationship between mental acuity, loneliness and depression, and social engagement among older adults. Specifically, late-life dementia has been associated with extended periods of loneliness in older adults\cite{wilson_loneliness_2007}. Another cognition study\cite{devanand1996depressed}, conducted a longitudinal study of adults aged 60 years or older living in North Manhattan, New York, and who were randomly selected from a dementia registry. Their study assessed the association between depressed mood and the onset of dementia. Physicians collected neuropsychological data to assess the degree of decreased cognitive function and determine the risk of dementia. Study results indicated that of the 1,070 participants, 218 (20\%) met the criteria for dementia at baseline assessment. Among the 852 participants that did not have dementia, depressive symptoms were common among those with cognitive impairment. Two years after the baseline data collection, follow-up data were collected on 478 participants who did not have dementia at baseline. A comparison of baseline and follow-up results concluded that of the 478 participants (93\%), the depressed mood was associated with dementia and exhibited symptoms of Alzheimer's disease \cite{devanand1996depressed}.

Before the turn of the last century, the only way to ascertain if a person has AD was via posthumous autopsy. Currently, as per the National Institute of Health (NIH), medical professionals ask the patient and their caregivers about overall health, medications, diet, medical history, and changes in behavior and personality. They may also administer a psychiatric evaluation to determine confounding causes and conduct tests on memory, problem-solving, attention, counting, language, blood, urine, and other standard medical tests. Finally, performing computed tomography (CT), magnetic resonance imaging (MRI), or positron emission tomography (PET) supports an AD diagnosis or rules out other plausible causes \cite{noauthor_how_nodate}. While there are other methods, such as accumulation of amyloid plaques and associated genes, these methods may not be entirely accurate \cite{thomas_objective_2020} \cite{giri_genes_2016}. Nonetheless, all methods listed are cost-prohibitive or require at least one dedicated medical professional. Consequently, researchers have been studying and modeling non-invasive methods using speech and linguistic features that do not necessitate human intervention to detect and evaluate AD patients. In addition, caregivers experience feelings of depression and being overwhelmed when caring for an older adult lacking social support mechanisms and are predominantly female and overwhelmingly low-income \cite{wilson_loneliness_2007}. 

Thus, with an aging world population negatively impacted by the symptoms associated with cognitive decline and an overwhelmed caregiving profession, research into technologies to help alleviate these issues is necessary. As AD affects the acoustics of speech\cite{rudzicz2012torgo} and vocabulary \cite{ferrer_growth_1983}, natural language processing and machine learning provide promising techniques for reliably detecting AD. While significant work has been done on detecting AD, this paper will evaluate and score mental status with ten different linear regression models using a combination of handcrafted or learned acoustic-linguistic features. The statistical significance and relevance of each selected feature are also studied. 

The rest of the paper covers a review of related works in Section 2. The models, dataset, feature extraction, feature selection, and training-testing protocol are detailed in Section 3. The performance of our models and features are compared to a state-of-the-art baseline linear model in section 4. The final section outlines the conclusion and future work.
\section{Related Works}

There has been significant research into the symptoms and manifestations of Alzheimer's Disease (AD) in medical literature and AD detection in interdisciplinary research. The review of relevant literature will be divided into two subsections: the first will cover the well-known acoustic-lingual expression of AD in patients, and the second will cover models and techniques currently used for evaluating and detecting AD. Furthermore, the first subsection helps establish the relevance of acoustic and linguistic features for AD progression, whereas the second subsection supports the reasoning behind our methodology.

\subsection{Acoustic and Linguistic Features in AD}
The relation between loss of memory and AD-associated neurodegeneration is well established. Recent research has studied acoustic and verbal aberrations present in patients with AD.
In particular, dysarthria/slurring, stuttering, monotony, higher delay, and associated acoustic features with AD \cite{ferrer_growth_1983}. Additionally, linguistic features such as paucity of words or aggramatism are also present with AD \cite{rudzicz2012torgo, lira2014analysis}. In severe cases, sentences uttered may comprise only nouns; articles, auxiliary verbs, and inflectional affixes are absent or replaced in lesser forms. Unsurprisingly, multiple approaches have utilized acoustic and linguistic features for the automatic detection of AD. We will discuss a few of these approaches in the following subsection.

\subsection{Contemporary Models and Techniques for AD Evaluation}

Speech has been used to distinguish between healthy and AD patients \cite{pulido_alzheimers_2020}. Some researchers have focused on developing dedicated machine learning model architectures \cite{chen2019attention, chien_assessment_2018, liu_new_2020} while others have focused on language models to classify AD \cite{guo_detecting_2019}. Some research has been focused on extracting acoustic and textual features that capture information indicative of AD, such as the length of segments and the amount of silence \cite{guo_detecting_2019}. Other researchers have used linguistic and audio features extracted from English speech \cite{fraser_linguistic_2016, gosztolya_identifying_2019}.
Prosodic features have been extracted from English speech \cite{nagumo_automatic_2020, qiao_computer_assisted_2020, ossewaarde_classification_2019} and so have paralinguistic acoustic features \cite{haider_assessment_2020}. Other approaches have attempted to focus on collecting speech from people performing multiple normative tasks to improve generalizability \cite{balagopalan_effect_2018}. However, most of these approaches utilize unbalanced, non-standardized, and proprietary datasets, which hampers their reproducibility and generalizability. We suggest the reader peruse this survey\cite{de2020artificial} to get a better understanding of these approaches.

 In 2020, The ADReSS Challenge \cite{luz2020alzheimer} defined shared tasks and standardized datasets with predefined metrics. Different approaches to the automated recognition of AD based on spontaneous speech and transcripts can be compared with two different tasks: AD Classification (AD vs. not-AD) and the neuropsychological score regression task. Furthermore, the challenge provided a baseline using standard machine learning models such as Random Forest and k-Nearest Neighbors on classification metrics (accuracy, precision, recall, F-1) and regression Root Mean Square Error (RMSE) scores. More details pertaining to the dataset are discussed in the Methodology section.
 
 Since the release of the dataset, significant work has been done on the classification task \cite{edwards2020multiscale, yuan2020disfluencies, pompili_inesc_id_2020}, the regression task \cite{farzana2020exploring}, or both \cite{balagopalan2020bert, syed2020automated, searle2020comparing, souganciouglu2020everything}. Of the two tasks, a high degree of accuracy 83\% to 92.84\% has been obtained on the classification task. However, the regression task, being the more challenging of the two, still has room for improvement and is the focus of this paper. Of the approaches reviewed, the lowest RMSE score of 4.56 was acheived on both training and testing sets and utilizes a linear Ridge Regressor model on a set of the 30 best correlating features \cite{balagopalan2020bert}. We refer to this work as the baseline and state-of-art for the comparison of our model and feature set through the remainder of the paper.
 
 \section{Methodology}
The models, dataset, feature extraction, feature selection, and training-testing protocol are detailed in the following subsections. All of the tasks performed were performed on a standard personal laptop machine or a Google Collaboratory notebook \cite{bisong_google_2019}. No specific accelerators are required, however, feature extraction, feature selection, and training-testing could be sped up through the utilization of more computing cores.

\subsection{The ADReSS Dataset and Metrics}
To enable comparison with the baseline, the ADReSS Challenge dataset \cite{luz2020alzheimer} is utilized. This dataset comprises of audio recordings, transcripts from patients performing the Cookie Theft task from the Boston Diagnostic Aphasia exam \cite{macwhinney2000childes}. Also provided with the dataset are metadata relating to the subject's age, gender and Mini Mental Status Examination (MMSE) score for both non-AD and AD patients. The regression task for this paper is associated with predicting these MMSE score based on the given audio recording and transcripts. Although the MMSE was originally designed to screen for dementia, it is an instrument currently used extensively to assess cognitive status in clinical settings \cite{wood2006assessing}. According to the Alzheimer's Association (2020), an MMSE score of 20–24 corresponds to mild dementia, 13–20 corresponds to moderate dementia, and a score < 12 is severe dementia.

Furthermore, the dataset comes divided into a Train Set (108 patients - 54 non-AD and 54 AD) and a Test Set (48 patients - 24 non-AD and 24 AD). As per the original challenge's guidelines and our baseline, the RMSE is used to determine and compare the performance of our approach. Since the dataset comes with many-to-one mapping of audio file to transcript files, in contrast to previous work, we opted to consider each unique audio-transcript file pair as a distinct observation. While this approach does limit us to shorter audio files with few utterances per file, the number of observations increases to 1447 for training and 569 for testing.

\subsection {Modeling and Train-Validation-Test Protocol}
Although the we were able to increase the sample size by considering audio-transcript file pairs, the number is still smaller than is demanded by most deep learning methods. While work such as \cite{keshari2020unravelling} has been done on small sample learning, these methods are still a black box. Interpretability is required to evaluate the association between features and the output of the model. While conventional, non-linear machine learning models such as Random Forest and k-Nearest Neighbors were originally the benchmark provided with the dataset \cite{luz2020alzheimer}, they have been outperformed by the baseline's linear models \cite{balagopalan2020bert} likely owing to the small sample size. Thus, we also opt for linear modeling. Similar to \cite{balagopalan2020bert}, we use regression models with in-built regularization or specific optimizations namely Ridge \cite{marquardt1975ridge}. Additionally, we also employ Lasso \cite{ranstam2018lasso}, ElasticNet \cite{zou2005regularization}, LassoLars \cite{tibshirani2011solution}, Bayesian Ridge \cite{bishop2006pattern}, Bayesian Automatic Relevance Determination \cite{wipf2007new}, Orthogonal Matching Pursuit \cite{blumensath2007difference}, Huber \cite{owen2006robust}, TheilSen \cite{wang2009theil}, and Stochastic Gradient Descent optimization \cite{bottou2012stochastic}. The models were trained and evaluated using a combination of the BSD-licensed scikit-learn \cite{scikit-learn}, numpy \cite{harris2020array}, seaborn \cite{Waskom2021}, scipy \cite{2020SciPy_NMeth}, and pandas \cite{reback2020pandas} package, and the PSF-licensed matplotlib \cite{Hunter_2007}. The ISF-licensed regressors \cite{haas_regressors_nodate} was used to evaluate the statiscal signficance of each selected feature . Beyond the default, the hyperparamters for each model can be found through the Appendix.

The training and testing protocol utilizes the provided disjoint sets provided with the dataset. Similar to the baseline, each model is trained using Leave One Subject Out (LOSO) Cross Validation on the training set and the RMSE is evaluated on both the training and test set. Of the models, Ridge, Lasso, ElasticNet, LassoLars, and Orthogonal Matching Pursuit's L1 or L2 regularization parameters were evaluated during this cross-validation. Additionally, a random 80-20 train-validation split of only the training set is used for feature selection.

\subsection {Feature Extraction, Pre-processing, and Feature Selection}

\subsubsection{Feature Extraction}
To learn from both the audio recording and text transcripts, feature extraction is necessitated. The dataset provides audio broken up into normalized audio chunks of the subject's sentences/utterances. And each participant's transcript only containing their text was combined into one text separated by a new line and was used for text feature extraction. To aid in our feature extraction a combination of software, and python libraries was used. Each of these third-party software, libraries, and their associated licenses are detailed in the Appendix. 

We further classify each feature into Audio Features and Linguistic Features. Each of these features may also either be handcrafted or learned. In total, each audio-transcript pair produced just over 13,000 features. To the best of our knowledge, a significant subset of these features are novel applications for the current task of MMSE score prediction.

* \textbf{Audio Features} (11,659 Features):

The learned audio features derived from audio recordings include Articulation \cite{vasquez2018towards,orozco2018neurospeech}, Phonation \cite{vasquez2018towards,arias2017parkinson}, and Prosody \cite{vasquez2018towards,dehak2007modeling} Features. Articulation features are made up of Bark band energies. Phonation features are composed up of pitch perturbation quotient, logarithmic energy, and derivatives of fundamental frequencies account for 28 features. Prosody features, based on energy and duration, include 103 features. The handcrafted audio features include spectral, Mel Frequency Cepstral Coefficients (MFCCs), and Chroma Vector/Deviation features. 
While all together these features total to 138, we utilized 80 different combinations of frame sizes and overlaps when the average feature are calculated. This was done to find the optimal frame size and overlap which would provide the most significant association with the given task during feature selection. 

* \textbf{Linguistic Features} (1,693 Features)
Linguistic features include, but are not limited to, Word/Sentence Count, Vocab Set, reading scales, and emotion analysis. These features were all extracted from the textual transcript files and totaled up to 1,693 features.

\subsubsection{Pre-processing}
Since audio data was retrieved from a normalized chunks no further pre-processing was required beyond feature extraction. Each participant's transcript was parsed and combined into one large string separated by a new line characters which was used for linguistic feature extraction. Lacking previous background and for convenient modeling, the features were scaled by the maximum value. The scaled features were normalized as required by the modeling library before training. No other pre-processing was performed.

\subsubsection{Feature Selection}

While extracting over 13,000 features provide us with a significant amount of data. Linear models, even with strong regularization, tend to get over-parameterized at this scale and require specific adaptation. Thus, we opt to select a subset of 100 features that provide the best estimation of the output. We selected 100 features due to limitations in available computing power and time. We utilized two methods from \cite{scikit-learn} for selecting the best features for this problem: (1) Recursive Feature Elimination and (2) Correlation Scores. For the first method, the best set of features which decreased the RMSE on a standard linear regression model trained on 80\% of the training set and minimized RMSE on the 20\% validation set was used. We could not get to a 100 features since the method only lets us select a minimum number of features required and outputted a set of features > 100. For the second method, we simply selected a the top 100 most correlated features with the output. In order, to further simply the model we trained and validated the models on features from the top 2 features until the all top features selected by the algorithms. Plots of validation RMSE for each of the methods can be seen in Figure 1. As expected, the error does incrementally decrease with the addition of each features. However, we are better suited taking a cut off around at a few feature after the steep decrease in RMSE. We chose to set this limit at 54 features which is half the number of subjects in the training set. Lacking precedence, we used P-values < 0.05 and coefficient > 0.01 were considered significant. Given page limitations, model summaries, source code, and additional plots are provided via the Appendix.
In the following section, we will cover the results of our modeling experiments and perform comparisons with the baseline.

\begin{figure}[t]
  \includegraphics[scale=0.70]{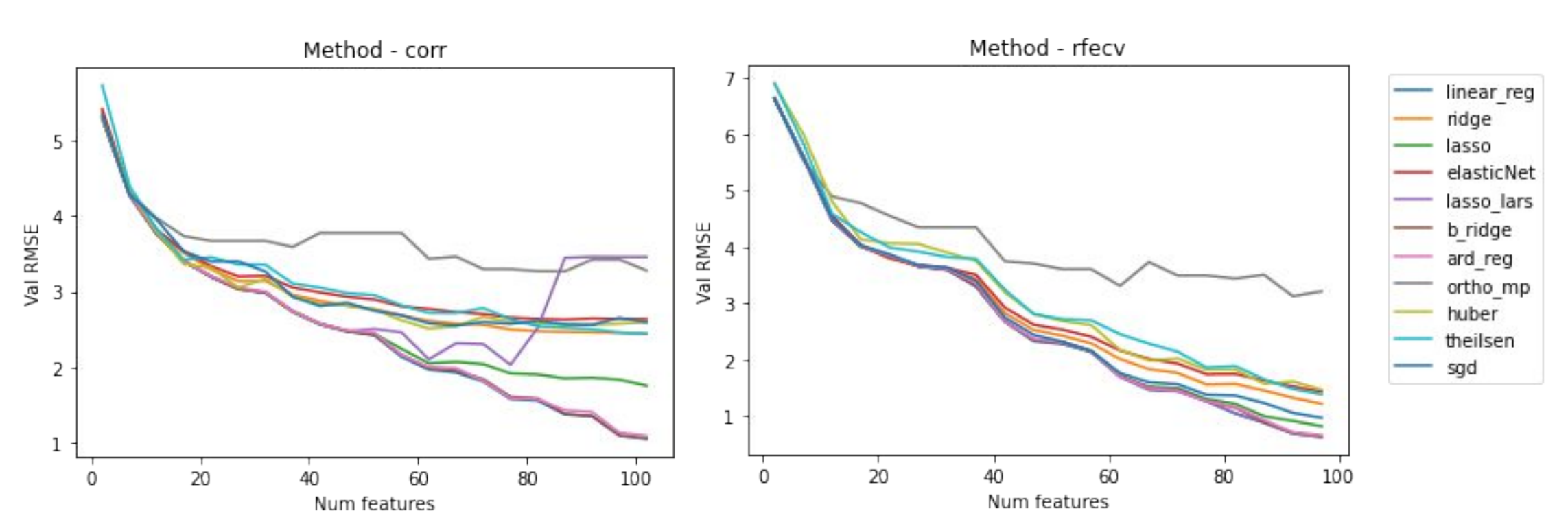}
  \centering
  \caption{Validation RMSE vs Num Features using Correlation and Recursive Elimination}
\end{figure}

\section{Results}

All of the models using features selected by both RFECV and Correlation outperformed the baseline model on the training set. Of these models, the standard linear regression model performed the best with an RMSE improvement of 2.37 compared to the baseline of 4.56. The RMSE plot for each model can be seen in Figure 2.

\begin{figure}[t]
  \includegraphics[scale=0.45]{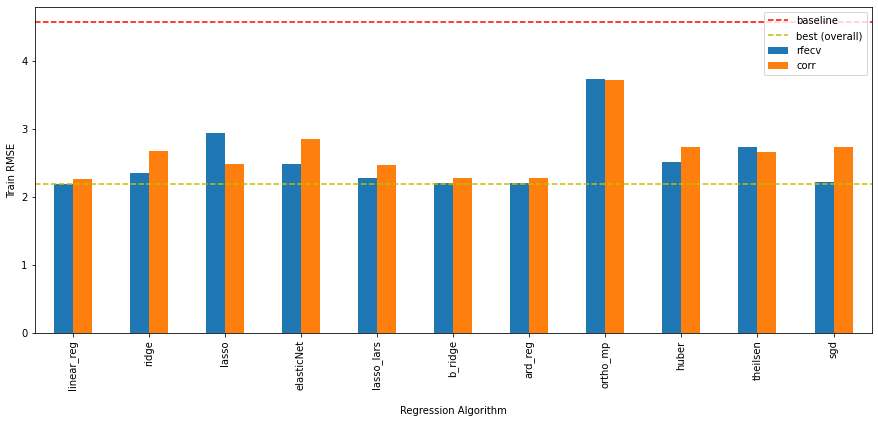}
  \centering
  \caption{Train RMSE for each model and each feature selection method}
\end{figure}

However, for the test set, not all models outperformed the baseline. Interestingly, none of the models which used features selected by recursive elimination outperformed the baseline whereas five models using correlation features outperformed the baseline despite the two methods having an overlap of 17 features selected out of the total 54. Of these models that outperformed the baseline, the stochastic gradient descent optimized model performed the best with an RMSE improvement of 0.66 compared to the baseline RMSE of 4.56. The plot of RMSE can be seen in Figure 3. 
\begin{figure}[t]
  \includegraphics[scale=0.44]{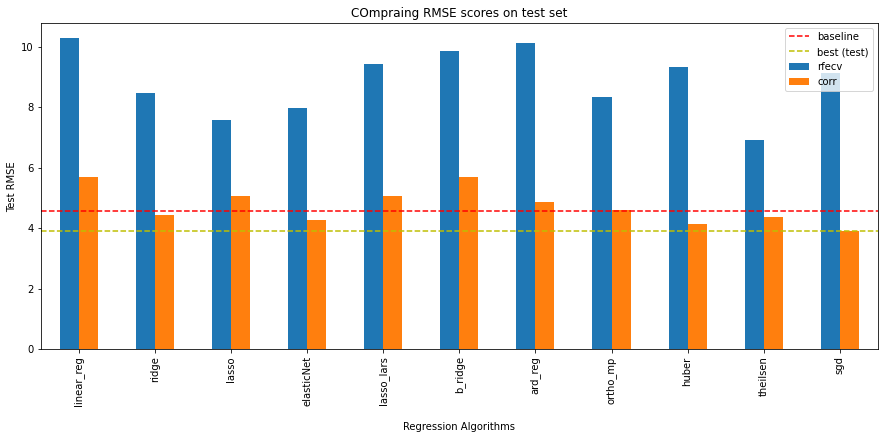}
  \centering
  \caption{Test RMSE for each model and each feature selection method}
\end{figure}

\begin{figure}[t]
  \includegraphics[scale=0.50]{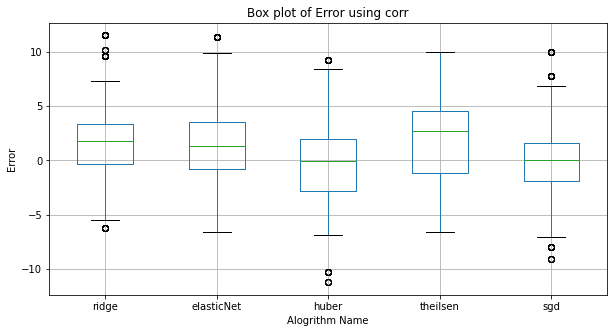}
  \centering
  \caption{Test RMSE for each model and each feature selection method}
\end{figure}

\begin{figure}[t]
  \includegraphics[scale=0.26]{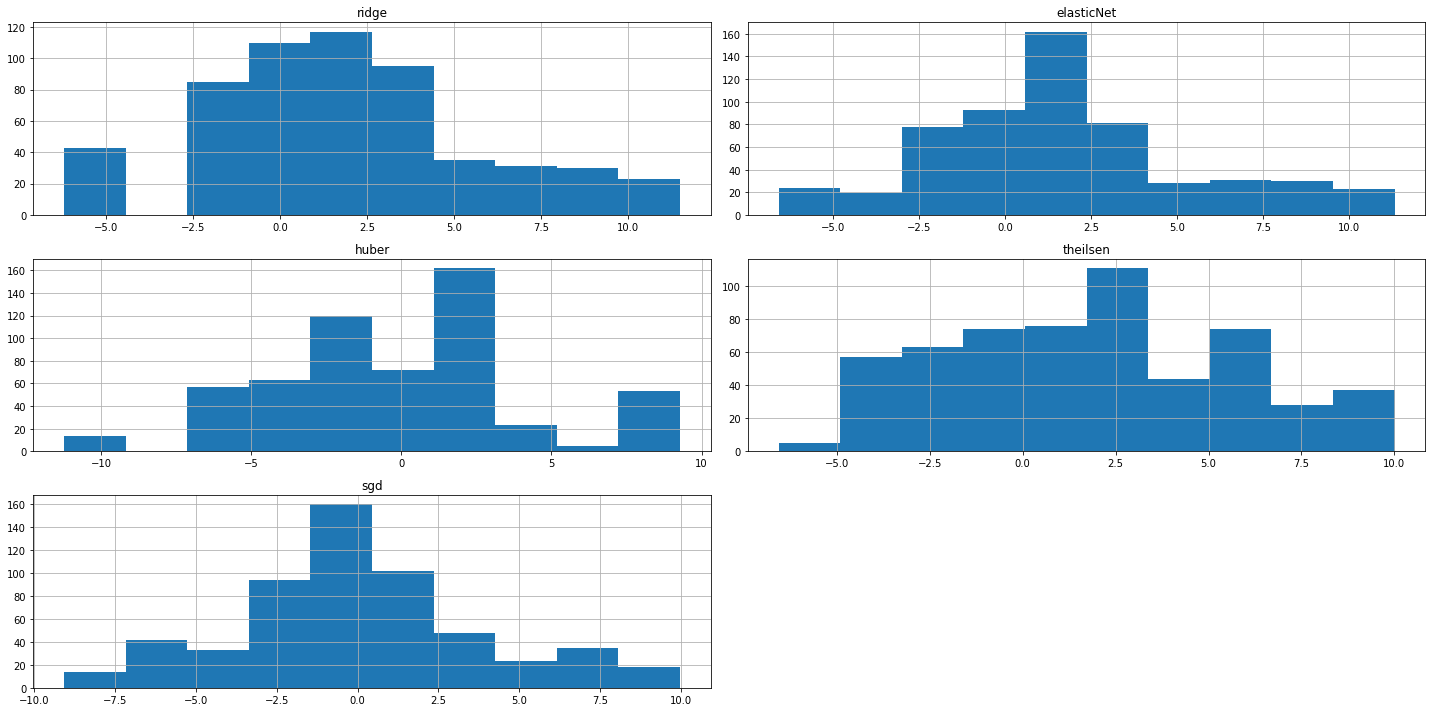}
  \centering
  \caption{Test RMSE for each model and each feature selection method}
\end{figure}
Upon a closer look into the the box in Figure 4 and histogram plots in Figure 5 of the residuals of each of the models that outperformed the baseline, we notice that stochastic gradient descent optimization has the most reliable performance on the test set. However, the range of prediction is still too large and unreliable in all of these models for real world application as of yet.  

Moreover, of the 54 features selected by the methods, it was noticed that all were handcrafted linguistic features related to word usage, readability, and character frequencies. This observation is inline with the observations of both the baseline and speech pathological research\cite{rudzicz2012torgo, lira2014analysis} that linguistic features are better predictors for this task in comparison to acoustic features and is supported. Details results of feature selection can be found via the Appendix

\section{Limitations and Future Work}
The major limitation of this work stems from data source. Since the dataset consists of audio recordings of the participants performing a specific task, it is unlikely these findings may be generalizable to recordings that are not obtained from the same task or for non-native English speakers. Furthermore, the standardization based on this task might also explain the proclivity of models to find significance of linguistic features over acoustic features for  the prediction of MMSE scores. It is possible that other modes of data capture may be better suited to a general approach for evaluating AD patients \cite{balagopalan_effect_2018}. 

Although the current dataset is remarkable, the sample size limits researchers from fully realizing and utilizing the most recent advancement in machine learning. While approaches such as early stopping and dropouts could be utilizes, one must question the external validity of such approaches within such a small sample size. Perhaps research into small sample size algorithms \cite{keshari2020unravelling} could be applied; however the issues related to interpretability still persists.

Contemporary research has shown the continued need to advance further the study of aging-associated disease effects on cognitive impairment in older adults\cite{li_aging_2021}. Researchers studied older adults who were already enrolled in research projects investigating the onset of Alzheimer's Disease (AD) on cognition under the assumption that the Functional Activities Questionnaire (FAQ) using the Instrumental Activities of Daily Living (IADL) scale to detect and track diminishing capability in managing and remembering daily household tasks and personal responsibilities. Difficulties in managing IADL identified in the FAQ proved helpful in detecting and tracking changes in cognition in healthy older adults at risk for Alzheimer's Disease \cite{a_marshall_functional_2015}. Furthermore, social determinants of health such as transportation, education, diet, and other daily factors negatively impact a person's health outlook. Black and Brown persons in the United States are adversely affected by schooling, diet, and disease symptoms associated with hypertension and diabetes that might cause cognitive decline \cite{landsberg2006therapeutic}. To further improvement the reliability of the models, perhaps social determinants, facial features, depression, and other correlates can be considered in junction with an in-home monitoring and audio-video capture device.

While we do believe that this paper sufficiently advance the state-of-the-art for this task, explores the largest feature space to date, and guides us towards automating the diagnosis of AD and modeling of cognitive status in the elderly, we must note that with automation we should not intend to replace trained medical professionals. We firmly believe that any technology stemming from research should be used as a tool to guide, assist, and ease medical professionals and caregivers to provide the best care possible.

\section{Conclusion}
While we were able to outperform the baseline with 5 different models, the performance of these models are still not fully suited for real world application. More research needs to be done to find models that work on low resource problems such as neurological evaluation of AD patients using audio and textual features.

\section{Acknowledgement}
The authors would like to acknowledge that the project was funded through a Research Award awarded through Amazon inc. We would also like to thank Mr. Bradon Thymes for helping with testing the installation of the DisVoice python library for acoustic feature extraction.

\bibliographystyle{ws-procs11x85}
\bibliography{refs}

\appendix

All supplemental materials can be found in the link below:
\href{https://bit.ly/3Skbaij}{https://bit.ly/3Skbaij}

\end{document}


\appendix

\section{Appendix}



\begin{table} [H]
  \caption{Models}
  \centering
  \begin{tabular}{llllll}
    \toprule
    Model     & Non Default Parameters\\
    \midrule
    Linear Regression & normalize=True      \\
    Ridge CV     & normalize=True     \\
    Lasso CV     & normalize=True,  cv=LeaveOneSubjectOut()     \\
    ElasticNet CV & normalize=True,  cv=LeaveOneSubjectOut()     \\
    LassoLars CV     & cv=LeaveOneSubjectOut(), njobs=-1, normalize=True      \\
    BayesianRidge     & normalize=True        \\
    ARD Regression & normalize=True       \\
    Orthogonal Matching Pursuit CV     & cv=LeaveOneSubjectOut(), njobs=-1, normalize=True       \\
    
    TheilSenRegressor    & None      \\
    SGDRegressor   & penalty=elasticnet, learning rate=adaptive        \\
    \bottomrule
  \end{tabular}
\end{table}

\begin{table}[H]
  \caption{Models RMSE}
  \centering
  \[
  \resizebox{\columnwidth}{!}{%
  \begin{tabular}{llllll}
    \toprule
    Model     & RFECV Train RMSE     & RFECV Test RMSE & Corr Train RMSE  & Corr Test RMSE\\
    \midrule
    Linear Regression & 2.1884  & 10.2744 & 2.2612 & 5.7028   \\
    Ridge     & 2.3464  & 8.4680   & 2.6666 & 4.4346    \\
    Lasso     & 2.9342     & 7.5849  & 2.4804 & 5.0565 \\
    ElasticNet & 2.4758  & 7.9799    & 2.8455 & 4.2832 \\
    LassoLars     & 2.2792 & 9.4232   & 2.4742 & 5.0488  \\
    BayesianRidge     & 2.1972  & 9.8555  & 2.2777 & 5.6853 \\
    ARD Regression & 2.2017  & 10.1135    & 2.2797 & 4.8767 \\
    Orthogonal Matching Pursuit &  3.7269   & 8.3418  & 3.7123 & 4.6058 \\
    Huber Regressor     &  2.5129    & 9.3455  & 2.7263 & 4.1282        \\
    TheilSenRegressor    & 2.7249 & 6.9160   & 2.6511 &  4.3723  \\
    SGDRegressor   & 2.2137 & 9.1226  & 2.7285 & 3.8958 \\
    \bottomrule
  \end{tabular}%
 }
 \]
\end{table}

\begin{table}[H]
  \centering

  \label{sample-table}
  
  \begin{tabular}{llllll}
    \toprule
    Model     & adjusted R\textsuperscript{2}     & F-statistic & Num of significant features\\
    \midrule
    Linear Regression & N/A & N/A & N/A   \\
    Ridge     & 0.8731 & 186.70   & 11  \\
    Lasso     & 0.8901  &  219.80  & 14 \\
    ElasticNet & 0.8565 &  160.82  & 8 \\
    LassoLars     & 0.89151  &  221.04  & 16 \\
    BayesianRidge     & 0.9081  & 265.44  & 24\\
    ARD Regression &  0.9079 &  264.95   &  29 \\
    Orthogonal Matching Pursuit &  0.7558   & 83.86  & 2 \\
    Huber Regressor   & 0.8683  &  177.50    &  17       \\
    TheilSenRegressor    & 0.8754  &  189.19  & 24  \\
    SGDRegressor   & N/A & N/A & N/A \\
    \bottomrule
  \end{tabular}
    \caption{Models Stats On Correlation}
\end{table}

\begin{table}[H]
  \caption{Models Stats On RFECV}
  \centering
  \begin{tabular}{llllll}
    \toprule
    Model     & adjusted R\textsuperscript{2}     & F-statistic & Num of significant features\\
    \midrule
    Linear Regression & 0.9151 &  289.71 & 39   \\
    Ridge     & 0.9024  &  248.63  &  30  \\
    Lasso     &  0.8474   & 149.72  & 17 \\
    ElasticNet & 0.8914  &  220.71  & 29 \\
    LassoLars     & 0.9079 &  265.08  & 31 \\
    BayesianRidge     & 0.9144  & 287.18  & 34\\
    ARD Regression &  0.9141 &  285.92   &  34 \\
    Orthogonal Matching Pursuit &  0.7538   & 83.00  & 5 \\
    Huber Regressor   & 0.8880  &  213.50    &  42       \\
    TheilSenRegressor    & 0.8684  &  177.71  & 32  \\
    SGDRegressor   & 0.9094 & 269.73 & 35 \\
    \bottomrule
  \end{tabular}
\end{table}

\begin{table}[H]
  \caption{Features}
  \centering
  \resizebox{\columnwidth}{!}{%
  \begin{tabular}{llllll}
    \toprule
    Feature Type     & Num Feature     & Extraction Method & Feature type \\
    \midrule
    Zero Crossing Rate     & 4   & pyAudioAnalysis\cite{giannakopoulos2015pyaudioanalysis} - Apache & HandCrafted Audio    \\
    Energy     & 4    & pyAudioAnalysis\cite{giannakopoulos2015pyaudioanalysis} - Apache & HandCrafted Audio      \\
    Entropy of Energy     & 4     & pyAudioAnalysis\cite{giannakopoulos2015pyaudioanalysis} - Apache  & HandCrafted Audio     \\
    Spectral Centroid     & 4    & pyAudioAnalysis\cite{giannakopoulos2015pyaudioanalysis} - Apache  & HandCrafted Audio    \\
    Spectral Spread     & 4     & pyAudioAnalysis\cite{giannakopoulos2015pyaudioanalysis} - Apache  & HandCrafted Audio      \\
    Spectral Entropy    & 4     & pyAudioAnalysis\cite{giannakopoulos2015pyaudioanalysis} - Apache  & HandCrafted Audio      \\
    Spectral Flux     & 4    & pyAudioAnalysis\cite{giannakopoulos2015pyaudioanalysis} - Apache  & HandCrafted Audio     \\
    Spectral Rolloff     & 4     & pyAudioAnalysis\cite{giannakopoulos2015pyaudioanalysis} - Apache  & HandCrafted Audio     \\
    MFCCs     & 48    & pyAudioAnalysis\cite{giannakopoulos2015pyaudioanalysis} - Apache  & HandCrafted Audio      \\
    Chroma Vector     & 44     & pyAudioAnalysis\cite{giannakopoulos2015pyaudioanalysis} - Apache  & HandCrafted Audio     \\
    Chroma Deviation     & 4     & pyAudioAnalysis\cite{giannakopoulos2015pyaudioanalysis} - Apache  & HandCrafted Audio     \\
    Articulation\cite{orozco2018neurospeech,vasquez2018towards}     & 489     & DisVoice - MIT & Learned Audio     \\
    Phonation\cite{vasquez2018towards,arias2017parkinson}    & 28     & DisVoice - MIT & Learned Audio      \\
    Prosody\cite{dehak2007modeling,vasquez2018towards}    & 103     & DisVoice - MIT & Learned Audio      \\
    Clausal and Phrasal complexity     & 379     & Taassc\cite{kyle2016measuring,lu2010automatic} - CC BY-NC 4.0 & Textual   \\
    Lexical Sophistication     & 840     & Taales\cite{kyle2018tool,kyle2015automatically} - CC BY-NC 4.0& Textual     \\
    Local and Global cohesion     & 168     & Taaco\cite{crossley2019tool,crossley2016tool} - CC BY-NC 4.0 & Textual     \\
    Text Numerical Analysis     & 28    & Sinlp\cite{crossley2014analyzing} - CC BY-NC 4.0 & Textual    \\
    Sentiment Analysis     & 271     & Seance\cite{crossley2017sentiment} - CC BY-NC 4.0 & Textual    \\
    Text Readability     & 9     & Arte\cite{crossley2022large} - CC BY-NC 4.0 & Textual \\
    Text Analysis     & 32     & Cla\cite{kyle2015native} - CC BY-NC 4.0 & Textual   \\
    \bottomrule
  \end{tabular}%
}
\end{table}

\bibliographystyle{ws-procs11x85}
\bibliography{refs}